\documentclass[12pt,twoside]{article}\usepackage{amsmath} 
\input BoxedEPS
\HideDisplacementBoxes
\usepackage{subfigure}
\SetRokickiEPSFSpecial
\usepackage{graphics}      
\usepackage{graphicx}      

\numberwithin{equation}{section}

\newcommand{\back}{\bar{g}_{\mu\nu}}
\newcommand{\lie}[1]{\mathcal{L}_{#1}}
\newcommand{\met}{g_{\mu\nu}}
\newcommand{\fluc}{h_{\mu\nu}}
\newcommand{\charge}[3]{Q_{#1} [#2, #3]}
\newcommand{\form}[3]{k_{#1} [#2, #3]}
\newcommand{\Eeq}{\mathcal{H}^{\mu\nu}}

\begin{document}
\title{Asymptotic Symmetries of Rindler Space at the Horizon and Null Infinity}\author{Hyeyoun Chung\\\small \it{Jefferson Physical Laboratory, Harvard University,}\\\small\it{17 Oxford St., Cambridge, MA 02138, USA}\\\small hyeyoun@physics.harvard.edu\\[-0.15in]} \date{\small\today} 


\maketitle

\begin{abstract}\noindent 


We investigate the asymptotic symmetries of Rindler space at null infinity and at the event horizon using both systematic and {\it ad hoc} methods. We find that the approaches that yield infinite-dimensional asymptotic symmetry algebras in the case of anti-de Sitter and flat spaces only give a finite-dimensional algebra for Rindler space at null infinity. We calculate the charges corresponding to these symmetries and confirm that they are finite, conserved, and integrable, and that the algebra of charges gives a representation of the asymptotic symmetry algebra. We also use relaxed boundary conditions to find infinite-dimensional asymptotic symmetry algebras for Rindler space at null infinity and at the event horizon. We compute the charges corresponding to these symmetries and confirm that they are finite and integrable. We also determine sufficient conditions for the charges to be conserved on-shell, and for the charge algebra to give a representation of the asymptotic symmetry algebra. In all cases, we find that the central extension of the charge algebra is trivial.
\end{abstract}

\section{Introduction}

Asymptotic symmetries have played an important role in defining conserved quantities even in purely classical theories of gravity\cite{Wald}. The Bondi and ADM energies of asymptotically flat spacetimes are charges corresponding to asymptotic time translations at the null and spatial infinities of flat space, respectively\cite{WaldText, BMS}, and Abbott and Deser showed that the Killing vectors of an arbitrary background spacetime $\back$ can be used to define conserved energy and momentum in a spacetime that asymptotes to $\back$ at infinity\cite{ADM}.

More recently, the theory of asymptotic symmetries has gained increased attention due to the growing importance of holographic dualities in physical theories\cite{WittenHolog}. The most important and well-known example of holographic duality is the AdS/CFT correspondence. This correspondence has its roots in the discovery of Brown and Henneaux that the asymptotic symmetry group of $\mathrm{AdS}_3$ is the two-dimensional conformal group\cite{BrownHenneaux}--a fact that led to the further insight that any consistent theory of quantum gravity on $\mathrm{AdS}_3$ is holographically dual to a two-dimensional conformal field theory.

This principle has been used to consider the possibility of holographic duals to gravitational theories on backgrounds other than AdS space\cite{BMS-CFT, WarpedAdS, Godel}. For example, the asymptotic symmetries of de Sitter space have been studied with a view towards developing a dS/CFT correspondence\cite{deSitter}. Once we have found the asymptotic symmetry group (ASG) of a spacetime, then we can hypothesize that the field theory dual to the theory of quantum gravity on that spacetime transforms under representations of the ASG.

The theory of asymptotic symmetries is of more than just academic interest: the insights given by the structure of the ASG and the asymptotic symmetry algebra appear to have real physical significance. For example, the Dirac bracket algebra of the ASG generators in $\mathrm{AdS}_3$ has a non-trivial central extension\cite{BrownHenneaux}. This fact was used to give a microscopic derivation of the Bekenstein-Hawking entropy for black holes with near horizon geometry that is locally $\mathrm{AdS}_3$\cite{Strominger}. It was later shown that the analysis used to obtain this result is very general, and applies to any consistent, unitary theory of quantum gravity if it contains the black hole as a classical solution. A similar approach has been used to derive the black hole entropy of an extreme four-dimensional Kerr black hole\cite{KerrCFT}, and the horizon entropy of deSitter space\cite{deSitter}.

Holography in flat space is still in the developmental stages. Although the asymptotic symmetry algebra for flat space has been derived in three and four dimensions\cite{BMS, Sachs, BBFlat}, we have not yet found a field theory that transforms under representations of the ASG. It is also an open problem to determine the ASG of Rindler space. (This problem was posed by Dionysios Anninos, who pointed out that while Rindler space is a subset of Minkowski space, this does not mean that their ASGs are identical.) The Unruh effect\cite{Birrell} shows that Rindler space can often have very different physical properties from Minkowski space. Also, the Einstein-Hilbert equations have interesting solutions that are asymptotically Rindler, such as the near-horizon geometry of the Schwarzschild black hole, which turns out to be Rindler$\times S^2$. We would therefore like to find the ASG of Rindler space at null infinity, and compare it to the ASG of flat space. Since Rindler space is bounded by past and future event horizons in addition to the boundary at infinity, we can also define an ASG of Rindler space at these boundaries. If the algebra of charges turns out to have a non-trivial central extension, we would like to see if the Cardy formula can be applied in the same way as in anti-de Sitter and de Sitter space to compute the horizon entropy of Rindler space\cite{Wilczek}.

This paper is structured as follows. In Section \ref{sec-AS} we review the formalism and definition of asymptotic symmetries, and describe a strategy for finding the ASG of a background spacetime. In Section \ref{sec-Rindler} we review the basic properties of Rindler space, and introduce the coordinates that we use in our analysis. We also describe some modifications to the usual formalism of asymptotic symmetries that must be made in order to apply the techniques to Rindler space. In Section \ref{sec-RedASG} we follow a systematic approach to try and find the ASG of Rindler space at null infinity. We then use a more {\it ad hoc} approach in an attempt to find an infinite-dimensional asymptotic symmetry algebra. In Section \ref{sec-Horizon} we study the asymptotic symmetries of Rindler space at the past event horizon. We conclude in Section \ref{sec-Discussion} with a summary and discussion of our results.

\section{Asymptotic Symmetries}\label{sec-AS}

In this section we describe the general concept of asymptotic symmetries, and outline two possible strategies for finding the asymptotic symmetry group of a spacetime. The formalism developed by Barnich and Brandt in \cite{BB} is particularly useful and elegant, and will be used throughout this paper.

We begin with a background spacetime with metric $\back$. A diffeomorphism generated by a vector $\xi$ will transform this metric to
\begin{equation}
\back + \lie{\xi}\back = \met = \back + \fluc
\end{equation}
for some perturbation $\fluc$. Defining the asymptotic symmetries of the background spacetime requires specifying both the allowed fluctuations $\fluc$, and the allowed diffeomorphisms $\xi$.

More precisely, the usual procedure for defining the ASG of the spacetime is as follows:
\begin{enumerate}
\item Define the {\it boundary} of the spacetime. In particular, it is necessary to define what it means to ``go to infinity.'' Usually this is done by taking a coordinate $r\to\infty$ while the other coordinates remain constant.
\item Define {\it boundary conditions} for the metric fluctuations $\fluc$. Usually this involves making some assumptions on the functional form of $\fluc$. In this work we assume that $\fluc$ can be expanded polynomially to at least second order in powers of $r$, as
\begin{equation}\label{eq-Expansion}
\fluc = \fluc^1r^m + \fluc^2r^{m-1} + o(r^{m-1})
\end{equation}
The boundary conditions specify $\fluc\to O(r^m)$ at infinity for some power $r^m$.
\item Find the most general diffeomorphisms $\xi$ that preserve the boundary conditions (i.e. satisfy $\lie{\xi}{\met} = O(\fluc)$ for any $\met$ satisfying the boundary conditions), and make sure that they form a well-defined algebra under the Lie bracket. These diffeomorphisms are the {\it candidate asymptotic symmetries}\cite{BBSchrod}. A subset of these will be promoted to elements of the ASG (and the corresponding algebra the asymptotic symmetry algebra) depending on whether or not they satisfy certain conditions, as described below. As with the perturbations $\fluc$, this step involves making assumptions about the form of $\xi$. We assume that $\xi$ can be expanded in powers of $r$, as $\xi^\mu = r^{m_\mu}\tilde{\xi}^\mu + o(r^{m_\mu})$, where $\tilde{\xi}^\mu$ is some function of coordinates other than $r$.
\item Calculate the {\it charges} associated with the candidate asymptotic symmetries. The formalism in \cite{BB} allows us to define the charges as surface integrals over the boundary $\partial\Sigma$ of a null or spacelike slice (the boundary need not be at infinity.) The charge corresponding to an asymptotic symmetry $\xi$ is given by
\begin{equation}\label{eq-Charge}
\charge{\xi}{h}{\bar{g}} = \int_{\partial\Sigma} \form{\xi}{h}{\bar{g}}
\end{equation}
where $k_\xi$ is an $n-2$ form constructed from the linearized equations of motion for $\fluc$, and $n$ is the number of spacetime dimensions). The explicit expression for $k_\xi$ is
\begin{eqnarray}\label{eq-Form}
\form{\xi}{h}{\bar{g}} &=& k_\xi^{[\nu\mu]}[h,\bar{g}](\mathrm{d}^{n-2}x)_{\nu\mu},\\
(\mathrm{d}^{n-p}x)_{\mu_1\ldots\mu_p}&:=&\frac{1}{p!(n-p)!}\epsilon_{\mu_1\ldots\mu_n}\mathrm{d}x^{\mu_{p+1}}\dots\mathrm{d}x^{\mu_n}\nonumber
\end{eqnarray}
where
\begin{align}
k_\xi^{[\nu\mu]}[h,\bar{g}] &= -\frac{\sqrt{-\bar{g}}}{16\pi}\biggl [\bar{D}^\nu(h\xi^\mu)+\bar{D}_\sigma(h^{\mu\sigma}\xi^\nu)+\bar{D}^\mu(h^{\nu\sigma}\xi_\sigma)\\
&\hspace{0.8cm}+\frac{3}{2}h\bar{D}^\mu\xi^\nu+\frac{3}{2}h^{\sigma\mu}\bar{D}^\nu\xi_\sigma+\frac{3}{2}h^{\nu\sigma}\bar{D}_\sigma\xi^{\mu}-(\mu \leftrightarrow \nu) \biggr]\nonumber
\end{align}
All indices are raised and lowered using the background metric. 

In general, this expression for $Q_\xi$ holds only for infinitesimal perturbations $\fluc$. For finite $\fluc$, we must integrate over a path $\gamma$ in phase space in order to compute the charges\cite{BBInteg}, which are given by
\begin{equation}
\int_\gamma D\delta g \int_{\partial\Sigma} k_\xi[\delta g, g(\gamma)]
\end{equation}
For consistency, the charge must be {\it integrable}, which means that it is independent of the path $\gamma$. A sufficient condition for integrability is
\begin{equation}
\int_{\partial\Sigma} \form{\xi}{\delta h_1}{g+\delta h_2} - \form{\xi}{\delta h_2}{g+\delta h_1} - \form{\xi}{\delta h_1 - \delta h_2}{g} = 0
\end{equation}
for any metric $\met$ allowed by the boundary conditions.

Integrability is also satisfied when a property called {\it asymptotic linearity} holds, which means that
\begin{equation}
\charge{\xi}{h}{\bar{g}} = \charge{\xi}{h}{\bar{g}+\delta g},
\end{equation}
or, in other words, the charge $Q_\xi$ has no non-linear corrections.

\item Determine whether the charges are {\it finite}, {\it integrable}, and {\it conserved on-shell}, where ``on-shell'' means that the perturbation $\fluc$ satisfies the linearized Einstein equations. The diffeomorphisms $\xi$ that correspond to non-zero charges satisfying these conditions are the elements of the ASG. The diffeomorphisms corresponding to trivial charges are trivial asymptotic symmetries. The ASG consists of all the allowed asymptotic symmetries that preserve the boundary conditions for $\fluc$ and correspond to finite, integrable, and conserved charges, modulo the trivial asymptotic symmetries. The vectors $\xi$ generating the diffeomorphisms are {\it asymptotic Killing vectors}.

\item Compute the Dirac bracket algebra of the charges. The algebra is defined by
\begin{eqnarray}\label{eq-Algebra}
\{Q_{\xi_1}, Q_{\xi_2}\} &:=& \delta_{\xi_2}Q_{\xi_1} = \charge{\xi_1}{\lie{\xi_2}g}{\bar{g}}
\end{eqnarray}
If certain technical conditions are satisfied, then the algebra of the charges provides a representation of the algebra of asymptotic symmetries $\xi$, up to a possible central extension $\mathcal{K}_{\xi_1, \xi_2}[\bar{g}]$, as shown below (where ``$\approx$'' indicates equality on-shell.)
\begin{eqnarray}
\{Q_{\xi_1}, Q_{\xi_2}\}&\approx& Q_{[\xi_1,\xi_2]} + \mathcal{K}_{\xi_1, \xi_2}[\bar{g}] - N_{[\xi_1,\xi_2]}[\bar{g}]
\end{eqnarray}
Here, $N_{[\xi_1,\xi_2]}[\bar{g}]$ is an arbitrary normalization constant that is usually set to zero. The explicit expression for the central charge $\mathcal{K}_{\xi_1, \xi_2}[\bar{g}]$ is 
\begin{equation}
\mathcal{K}_{\xi_1, \xi_2}[\bar{g}] = \charge{\xi_1}{\lie{\xi_2}\bar{g}}{\bar{g}}
\end{equation}
The central charge is non-trivial if it cannot be reabsorbed into $N_{[\xi_1,\xi_2]}[\bar{g}]$. When the background metric satisfies the Einstein equations $\bar{R}_{\mu\nu} = 2(n-2)^{-1}\Lambda\back$, the explicit expression for the central charge is
\begin{align}
\mathcal{K}_{\xi_1, \xi_2}[\bar{g}] &= \frac{1}{16\pi}\int_{\partial\Sigma} (\mathrm{d}^{n-2}x)_{\nu\mu}\sqrt{-\bar{g}}\biggl[-2\bar{D}_\rho\xi_1^{\rho}\bar{D}^\nu\xi_2^\mu+2\bar{D}_\rho\xi_2^\rho\bar{D}^\nu\xi_1^\mu\\
&\hspace{1.5cm}+4\bar{D}_\rho\xi_1^\nu\bar{D}^\rho\xi_2^\mu+(\bar{D}^\rho\xi_2^\nu+\bar{D}^\nu\xi_2^\rho)(\bar{D}^\mu\xi_{1\rho}+\bar{D}_\rho\xi_1^\mu)\nonumber\\
&\hspace{1.5cm}+\frac{8\Lambda}{2-n}\xi_1^\nu\xi_2^\mu+2\bar{R}^{\mu\nu\rho\sigma}\xi_{1\rho}\xi_{2\sigma}\biggr]\nonumber
\end{align}
\end{enumerate}
The boundary conditions are part of the specification of the ASG. In fact, different boundary conditions can give rise to different asymptotic symmetry groups. It is important to choose fall-offs for $\fluc$ that are relaxed enough to include physically interesting solutions, but restrictive enough so that the charges are well-defined.

\subsection{A Strategy for finding the ASG}\label{sec-Strategy}

The formalism in \cite{BB} provides sufficient conditions for the charges $Q_{\xi}$ to be finite and conserved. Specifically, if $\Eeq$ are the linearized Einstein equations, then the charges are finite and conserved on-shell if
\begin{equation}\label{eq-BBCond}
\lie{\xi}{\back}\Eeq\mathrm{d}^nx \to 0
\end{equation}
This condition suggests a more systematic approach for finding an ASG than simply guessing an appropriate set of boundary conditions\cite{BBAdS}. First, we find the exact Killing vectors $\xi_0$ of the background, and determine their fall-offs at infinity, which will be of the form $\xi_0^\mu \to O(r^{m_\mu})$. Then, for arbitrary vector fields $\xi$ that satisfy the fall-off conditions of the exact Killing vectors, we determine the fall-off of $\lie{\xi}{\back}$:
\begin{equation}
\lie{\xi}{\back} \to O(r^{m_{\mu\nu}}),
\end{equation}
and find the most general $\xi$ that satisfy the {\it asymptotic Killing equations} $\lie{\xi}{\back} \to o(r^{m_{\mu\nu}})$. This is the class of candidate asymptotic Killing vectors. For these $\xi$, we obtain a new set of fall-offs
\begin{equation}
\lie{\xi}{\back} \to O(r^{p_{\mu\nu}})
\end{equation}
Looking at the sufficient condition (\ref{eq-BBCond}) for finite and conserved charges, this gives us the following fall-off conditions on $\Eeq$:
\begin{equation}
\Eeq\mathrm{d}^nx \to o\left(\frac{1}{r^{p_{\mu\nu}}}\right)
\end{equation}
This in turn gives us boundary conditions on $\fluc$, as $\Eeq$ is given in terms of $\fluc$ by
\begin{align}\label{eq-Eeq}
\Eeq[h,\bar{g}] &=\frac{\sqrt{-\bar{g}}}{32\pi}\biggl[\frac{2\Lambda}{n-2}(2h^{\mu\nu}-\bar{g}^{\mu\nu}h)+\bar{D}^\mu\bar{D}^\nu h+\bar{D}^\lambda\bar{D}_\lambda h^{\mu\nu}\\
&\hspace{1.8cm}-2\bar{D}_\lambda\bar{D}^{(\mu}h^{\nu)\lambda}-\bar{g}^{\mu\nu}(\bar{D}^\lambda\bar{D}_\lambda h - \bar{D}_\lambda\bar{D}_\rho h^{\rho\lambda})\biggr]\nonumber
\end{align}
where all indices are raised and lowered using the background metric. Once we have determined boundary conditions for $\fluc$, we check to make sure that the candidate asymptotic Killing vectors $\xi$ preserve these boundary conditions, and form a well-defined algebra. Finally, we compute the charges corresponding to $\xi$, check that they are integrable, and calculate the Dirac bracket algebra and central extension.

Asymptotic Killing vectors that automatically satisfy the asymptotic Killing equations are deemed to be trivial. At the end of the analysis it is necessary to check that these vectors do indeed give rise to trivial charges, so that they also satisfy our original definition of trivial asymptotic Killing vectors.

In this paper we will use both of the approaches described above to try and determine the ASG of Rindler space. The second is more systematic, but as the conditions for conservation and finiteness of the charges are sufficient and not necessary, we find that this method can lead to a severely restricted ASG. Therefore, in order to find an infinite-dimensional asymptotic symmetry algebra we are forced to impose boundary conditions by hand, using intuition and trial-and-error. 

\section{Rindler Space}\label{sec-Rindler}

In this section we briefly review some properties of Rindler space\cite{Birrell}, and describe some of the modifications that we make to the Barnich-Brandt formalism in order to apply their techniques to Rindler space. 

\subsection{Rindler Coordinates}\label{sec-RindlerCoords}

\begin{figure}[ht]$$\BoxedEPSF{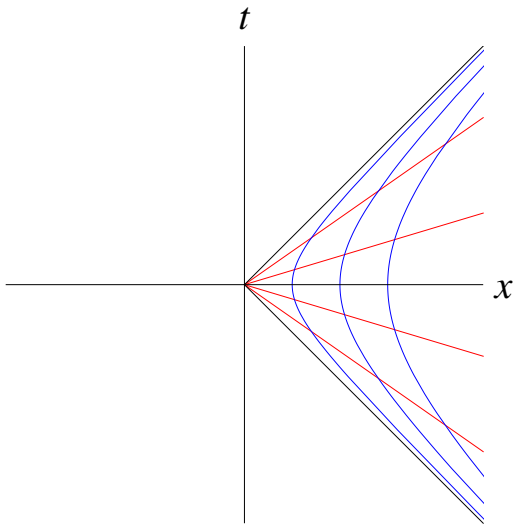 scaled 500}$$\caption{Lines of constant $\eta$ (red) and constant $\zeta$ (blue) in Rindler space.}\label{fig-Coords}\end{figure}

\begin{figure}[ht]$$\BoxedEPSF{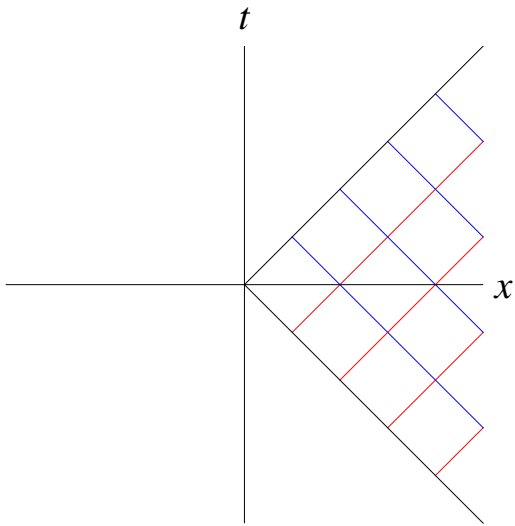 scaled 500}$$\caption{Lines of constant $u$ (red) and constant $r$ (blue) in Rindler space.}\label{fig-NullCoords}\end{figure}

In ordinary 4-dimensional Minkowski space, we can make a change of coordinates that is adapted to uniformly accelerated motion. Let $x$ be the the direction of acceleration, and make the coordinate transformation
\begin{eqnarray}
t &=& \zeta\sinh{a\eta}\\
x &=& \zeta\cosh{a\eta}\\
y &=& y\\
z &=& z
\end{eqnarray}
where $a$ is a positive constant and the new coordinates have ranges $-\infty < \eta < \infty$ and $0 < \zeta < \infty$. These coordinates cover the wedge $x > |t|$ of Minkowski space: this region is known as Rindler space. The metric in these new coordinates takes the form:
\begin{equation}
\mathrm{d}s^2 = -a^2\zeta^2\mathrm{d}\eta^2 + \mathrm{d}\zeta^2+\mathrm{d}y^2+\mathrm{d}z^2
\end{equation}
The lines of constant $\eta$ and constant $\zeta$ are shown in Figure \ref{fig-Coords}.
The lines of constant $\zeta$ are hyperbolae described by
\begin{equation}
x^2 - t^2 = \zeta^2,
\end{equation}
and are therefore worldlines of uniformly accelerated observers (also known as Rindler observers). An observer with constant acceleration of magnitude $\alpha$ in the $x$-direction and measuring proper time $\tau$ travels along the path
\begin{eqnarray}
\eta(\tau) &=& \frac{\alpha}{a}\tau\\
\zeta(\tau) &=& \frac{1}{\alpha}
\end{eqnarray}
In this work we make use of yet another set of coordinates for Rindler space in order to make our analysis simpler. First we make the coordinate transformation
\begin{eqnarray}
a\tilde{u} &=& a\eta - \ln{a\zeta}\\
a\tilde{r} &=& a\eta + \ln{a\zeta}.
\end{eqnarray}
followed by another coordinate transformation
\begin{eqnarray}
u &=& e^{-a\tilde{u}}\\
r &=& e^{a\tilde{r}}
\end{eqnarray}
Under this change of coordinates the metric becomes
\begin{eqnarray}\label{eq-URMetric}
\mathrm{d}s^2 = \frac{1}{a^2}\mathrm{d}u\mathrm{d}r +\mathrm{d}y^2+\mathrm{d}z^2
\end{eqnarray}
Rindler space corresponds to the ranges $0 < u,r < \infty$. The lines of constant $u$ and $r$ (shown in Figure \ref{fig-NullCoords}) correspond to the null lines in Rindler space: that is, the lines of constant $x+t$ and $x-t$.
We also make use of the coordinates $(v,r)$, where $v:=\tilde{u}$. With these coordinates, the metric becomes
\begin{eqnarray}\label{eq-vMetric}
\mathrm{d}s^2 = -\frac{e^{-av}}{a}\mathrm{d}v\mathrm{d}r +\mathrm{d}y^2+\mathrm{d}z^2
\end{eqnarray}
In this case Rindler space corresponds to the ranges $-\infty<v<\infty$, $0<r<\infty$.

\subsection{The Causal Structure of Rindler Space}\label{sec-Causal}

\begin{figure}[ht]$$\BoxedEPSF{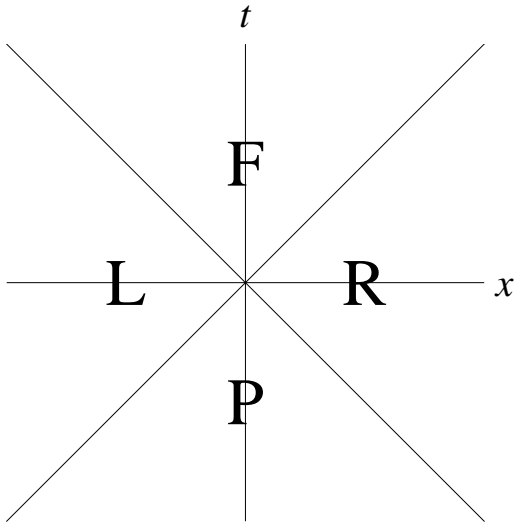 scaled 500}$$\caption{The division of flat space into the Right Rindler wedge (R), the Left Rindler wedge (L), and the future (F) and past (P) regions.}\label{fig-Wedges}\end{figure}

\begin{figure}[ht]$$\BoxedEPSF{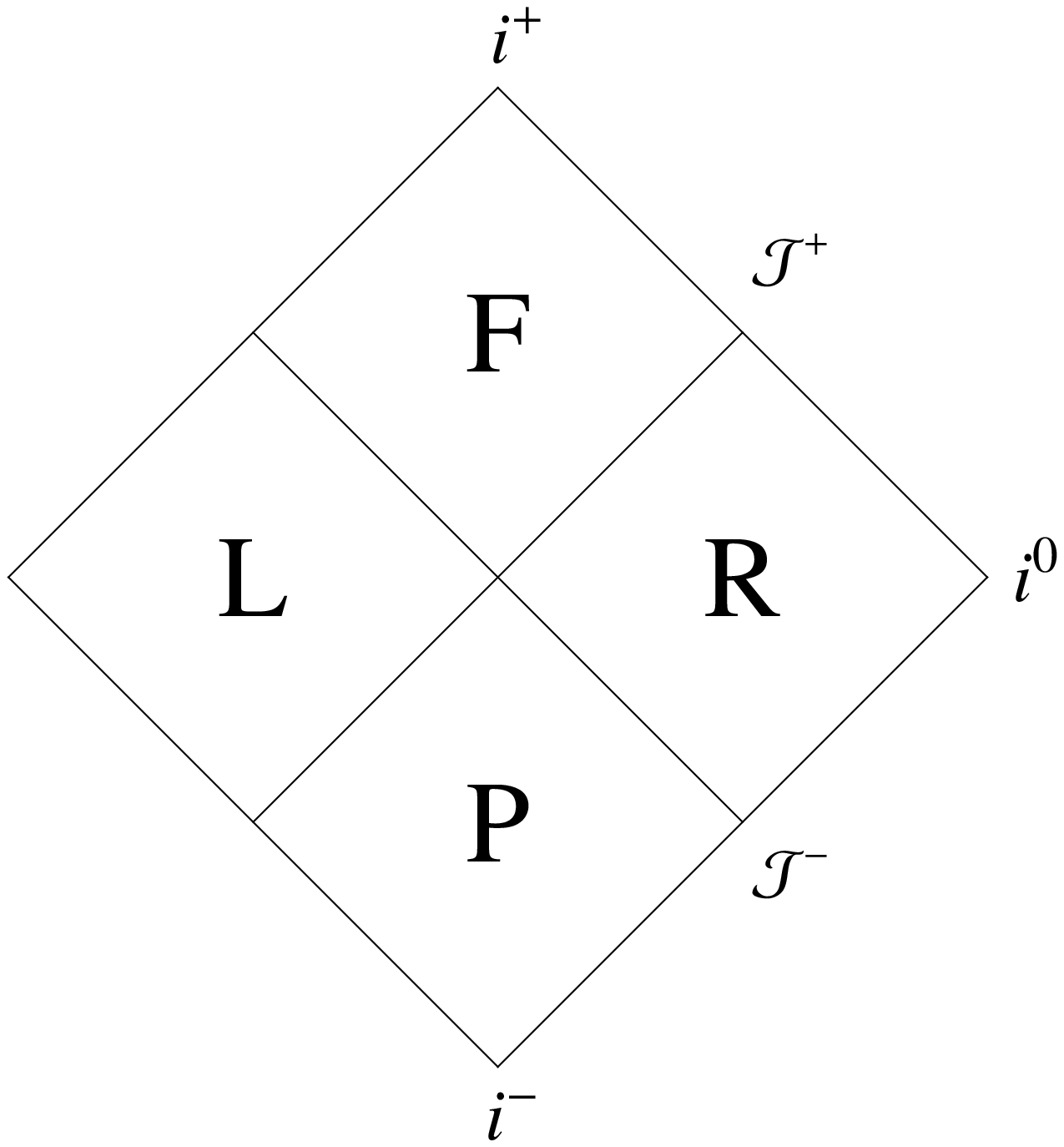 scaled 500}$$\caption{The Penrose diagram of Minkowski space, showing the regions R, L, F, and P. The conformal infinity of the Right Rindler wedge is a subset of the conformal infinity of Minkowski space, intersecting $\mathcal{J}^+$ and $\mathcal{J}^-$.}\label{fig-Penrose}\end{figure}

Rindler space covers only a wedge $x > |t|$ of Minkowski space, known as the Right Rindler wedge (R). We can also define the Left Rindler wedge (L) and the future (F) and past (P) regions, as shown in Figure \ref{fig-Wedges}.
Because Rindler observers approach but do not cross the null rays $u=0, r=0$, these rays act as event horizons. The regions R, L, F, and P also appear on the Penrose diagram for Minkowski space, shown in Figure \ref{fig-Penrose}.
Rindler observers emerge from $\mathcal{J}^-$ and eventually reach $\mathcal{J}^+$. Therefore, the infinity of Rindler space is a subset of the infinity of Minkowski space. 

Rindler space is bounded both by the event horizons and by the boundary at infinity. In this paper we investigate both boundaries, approaching the horizon along lines of contant $u$, $y$, and $z$, as $r\to 0$, and null infinity along lines of constant $v$, $y$, and $z$, as $r\to\infty$. The metric in Eq.(\ref{eq-URMetric}) is that of flat space, but this analysis differs from the study of the ASG of 3-dimensional flat space in \cite{BBFlat}. When looking at flat space we consider the entire spacetime, with $-\infty<t,x,y<\infty$. Therefore, in \cite{BBFlat} the charges are calculated on null slices of constant $\bar{u}:= t-\bar{r}$, where $\bar{r}:= \sqrt{x^2+y^2}$ is the usual radial coordinate, and the ``time'' coordinate $\bar{u}$ is allowed to take the range $-\infty<\bar{u}<\infty$. This means that when we define a general metric $g_{\mu\nu}$ as being ``asymptotically flat,'' we are requiring it to satisfy boundary conditions as $t+\bar{r}\to\infty$ for all real values of $\bar{u}$. i.e. we require the boundary conditions to be met on the whole of $\mathcal{J}^+$.

In this work we use the coordinate $v$, introduced in Section \ref{sec-RindlerCoords}, as our ``time'' coordinate. As $v$ ranges from $-\infty$ to $\infty$ we cover the region $x>t$ of flat space. If we then take the range $0<r<\infty$, we cover the right Rindler wedge. We approach $\mathcal{J}^+$ by taking $t+x\to\infty$ for constant $\ln(t-x)$, $y$, and $z$. Thus, when we define a general metric $g_{\mu\nu}$ as being ``asymptotically Rindler'' at infinity, we are requiring it to satisfy boundary conditions on the intersection of $\mathcal{J}^+$ with Rindler space. We therefore obtain different results from the study of the ASG of Minkowski space at null infinity.

\subsection{Modifications of the ASG Formalism}\label{sec-RindlerMod}

We will have to make two modifications to the Barnich-Brandt formalism in order to define charges in Rindler space. In most background spacetimes, it is possible to define a foliation of spacelike or null slices $\Sigma$ such that the entire boundary $\partial\Sigma$ of a slice is at infinity. The charges are then computed by integrating over $\partial\Sigma$. However, in Rindler space this is difficult to do in a natural way. Therefore, we will work with null slices defined by constant $u$ and covering the region $0<r<\infty$. When taking the limit $r\to\infty$, there are no constraints on the form of the asymptotic Killing vectors $\xi$ at $r\to 0$, and vice versa. Therefore we may consider the asymptotic symmetries and the charges at infinity separately from those at the horizon. We can also take the form of the vectors $\xi$ to be such that the contribution to the charges from the other parts of the boundary is zero, since the form of $\xi$ is completely unconstrained for $r\not\to 0$ and $r\not\to\infty$.

Secondly, the parts of the boundary at $r\to 0$ and $r\to\infty$ are infinite in the transverse $y$ and $z$ directions, so a regulator has to be introduced to make sure that the charges are finite\cite{BBSchrod}. We choose to integrate over a finite box in the $y$-$z$ plane. Since the action of an asymptotic symmetry $\xi$ can change the shape of the box of integration, we have to modify the definition of the Dirac bracket given in (\ref{eq-Algebra}) to take this change into account. We use the following definition of the modified Dirac bracket:
\begin{eqnarray}\label{eq-ModBracket}
\{Q_{\xi_1}^{\mathrm{box}}[h,\bar{g}], Q_{\xi_2}^{\mathrm{box}}[h,\bar{g}]\} &:=& \delta_{\xi_2}^{h}Q_{\xi_1}^{\mathrm{box}}[h,\bar{g}] + \delta_{\xi_2}^{\mathrm{box}}Q_{\xi_1}^{\mathrm{box}}[h,\bar{g}]
\end{eqnarray}
The first term is the usual Dirac bracket given by (\ref{eq-Algebra}), and the second term is defined to be
\begin{equation}
\delta_{\xi_2}^{\mathrm{box}}Q_{\xi_1}^{\mathrm{box}}[h,\bar{g}] := \int_{\partial\Sigma} \mathcal{L}_{\xi_2} k_{\xi_1}[h,\bar{g}].
\end{equation}
This takes into account the change in the domain of integration due to the diffeomorphism $\xi_2$.

\section{The Asymptotic Symmetry Group of Rindler Space at Null Infinity}\label{sec-RedASG}

In this section we investigate the ASG of Rindler space at null infinity, using the coordinates $(v,r,y,z)$. We first follow the strategy outlined in Section \ref{sec-Strategy} for finding the ASG. (The details of the calculation are given in Appendix \ref{app-Red}.) The components of the exact Killing vectors $\xi_0$ of the background fall off like
\begin{equation}
\xi_0^v \rightarrow O(1),\hspace{1cm}\xi_0^r,\,\xi_0^y,\,\xi_0^z \rightarrow O(r)
\end{equation}
at infinity. For arbitrary vectors $\xi$ with the same fall-offs as the exact Killing vectors, we find the following asymptotic Killing equations:
\begin{align}\label{eq-AKE1}
\lie{\xi}{\bar{g}_{vv}} &\rightarrow o(r),\hspace{1cm} \lie{\xi}{\bar{g}_{vr}} \rightarrow 0,\hspace{1cm} \lie{\xi}{\bar{g}_{vy}} \rightarrow o(r),\hspace{1cm} \lie{\xi}{\bar{g}_{vz}} \rightarrow o(r)\\
\lie{\xi}{\bar{g}_{rr}} &\rightarrow o(1/r),\hspace{0.6cm} \lie{\xi}{\bar{g}_{ry}} \rightarrow 0,\hspace{1cm}\label{eq-AKE2} \lie{\xi}{\bar{g}_{rz}} \rightarrow 0\\
\lie{\xi}{\bar{g}_{yy}} &\rightarrow o(r),\hspace{1cm} \lie{\xi}{\bar{g}_{yz}} \rightarrow o(r),\hspace{0.5cm} \lie{\xi}{\bar{g}_{zz}} \rightarrow o(r)\label{eq-AKE3}
\end{align}
Solving the Killing equations to leading order gives the following form for the candidate asymptotic Killing vectors:
\begin{eqnarray}
\xi^v &=& e^{av}(A_1 + A_2z + B_1y + C_1e^{-av}) + o(r^0)\\
\xi^r &=& aC_1r + o(r)\nonumber\\
\xi^y &=& \frac{B_1r}{2a} + o(r)\nonumber\\
\xi^z &=& \frac{A_2r}{2a} + o(r)\nonumber
\end{eqnarray}
for arbitrary constants $A_1, A_2, B_1,$ and $C_1$. Vectors of the form
\begin{equation}
\xi^v \rightarrow 0,\,\,\,\,\,\,\, \xi^r,\,\xi^y,\,\xi^z \rightarrow o(r)
\end{equation}
automatically satisfy the asymptotic Killing equations to leading order, and are therefore trivial. We already see that we obtain very different results from the case of three-dimensional AdS space\cite{BrownHenneaux, BBAdS} or even Minkowski space\cite{BBFlat, BB}, where the algebra of asymptotic symmetry vectors was infinite-dimensional. Here we have a finite-dimensional set of candidate asymptotic symmetries. The form of $\lie{\xi}{\back}$ at infinity determines the fall-offs for $\Eeq$ in order to obtain finite and conserved charges. Computing $\Eeq$ in terms of $\fluc$, we find the corresponding boundary conditions on $\fluc$:
\begin{align}
h_{\mu\nu} &= \left (\begin{array}{cccc} O(1) & O(1/r^2) & O(1/r) & O(1/r) \\
h_{rv} = h_{vr} & O(1/r^2) & O(1/r^2) & O(1/r^2)\\
h_{yv} = h_{vy} & h_{yr}=h_{ry} & O(1/r) & O(1/r)\\
h_{zv} = h_{vz} & h_{zr}=h_{rz} & h_{zy} = h_{yz} & O(1/r)
\end{array} \right )
\end{align}
We then impose consistency by requiring that the asymptotic Killing vectors preserve these boundary conditions, i.e. that $\lie{\xi}{\back} = O(\fluc)$. This requirement, together with the condition that the asymptotic Killing vectors should form a well-defined algebra under the Lie bracket, reduces the candidate asymptotic symmetries to vectors of the form
\begin{eqnarray}
\xi^v &=& e^{av}\left(A_1 + C_1e^{-av} + \frac{V_1}{r}\right)+O(1/r^2)\\
\xi^r &=& aC_1r + R_1 + O(1/r)\nonumber\\
\xi^y &=& Y_1 + O(1/r)\nonumber\\
\xi^z &=& Z_1 + O(1/r)\nonumber
\end{eqnarray}
for arbitrary constants $A_1, C_1, R_1, V_1, Y_1$, and $Z_1$. Modulo trivial asymptotic Killing vectors, the asymptotic symmetry algebra is generated by the vectors
\begin{eqnarray}
\xi_1 &=& e^{av}\partial_v\\
\xi_2 &=& \partial_v + ar\partial_r
\end{eqnarray}
The Lie bracket algebra of asymptotic symmetries is given by
\begin{align}\label{eq-LBAlg1}
[\xi_1, \xi_2] &= -a\xi_1\\
[\xi_1, \xi_1] &= [\xi_2, \xi_2] = 0
\end{align}
As a final step we want to compute the charges corresponding to these vectors. These charges are finite and conserved by construction, although we will still have to check that integrability holds. The charges are given by Eq.(\ref{eq-Charge}), where we integrate over the boundary of a slice $\Sigma$ of constant $v$. As described in Section \ref{sec-RindlerMod}, we integrate over a finite box in the $y$-$z$ plane and assume that the only contribution to the integral comes from the part of $\partial\Sigma$ at $r\to\infty$. As the asymptotic Killing vectors do not transform the box in the $y$-$z$ plane, we can write
\begin{equation}\label{eq-ChargeNullCoords}
Q_{\xi}[h,\bar{g}] = \lim_{r\to\infty} \int_{\partial\Sigma} \mathrm{d}y\,\mathrm{d}z\,\, k_{\xi}^{[vr]}[h,\bar{g}].
\end{equation}
where $k_{\xi}^{[vr]}$ is given by Eq.(\ref{eq-Form}). We find that:
\begin{eqnarray}
Q_{\xi_1}[h,\bar{g}] &=& 0\\
Q_{\xi_2}[h,\bar{g}] &=& 0
\end{eqnarray}
for $\fluc$ obeying the boundary conditions. Therefore finiteness, conservation, and integrability around $\back$ are automatically satisfied. The Dirac bracket algebra of the charges as defined by Eq.(\ref{eq-Algebra}) also gives a trivial representation of the Lie bracket algebra of asymptotic symmetries given in Eq.(\ref{eq-LBAlg1}). The central charge of the Dirac bracket algebra is also zero. The asymptotic Killing vectors that we defined to be trivial also give rise to zero charges, thus justifying our definition.

\subsection{Comparison to Flat Space}\label{sec-Flat}

As mentioned in Section \ref{sec-Causal}, we might have expected the ASG of Rindler space at null infinity to be similar to that of flat space, as Rindler space is just a subset of flat space, and they share a similar conformal structure at infinity. However, it turns out that their asymptotic symmetries are very different. In \cite{BBFlat} the asymptotic symmetries of flat space at null infinity were analyzed using the same method as we used above for Rindler space. Instead of finding a finite-dimensional algebra of asymptotic symmetry vectors, the authors found that they recovered the Bondi-Metzner-Sachs (BMS) algebra\cite{BMS}. The asymptotic Killing vectors generating the BMS algebra correspond to charges that represent the BMS algebra under the Dirac bracket, with a non-trivial central extension.

It is natural to speculate if it is in fact {\it possible} to find an infinite-dimensional ASG for Rindler space, and we have merely overlooked the appropriate boundary conditions that would lead to such a group by being overly strict in our methods. The Barnich-Brandt formalism gives sufficient, but not necessary, conditions to find a consistent set of $\fluc$ and $\xi$ defining an ASG. So it is reasonable to consider relaxing their conditions in order to find a set of candidate asymptotic symmetries, and then checking finiteness, conservation, and integrability of the corresponding charges by hand.

One option is to relax the conditions on some of the components of the asymptotic Killing equations. That is, if the asymptotic Killing equations require $\lie{\xi}{\back} \rightarrow o(r^{m_{\mu\nu}})$, then we can impose the condition $\lie{\xi}{\back} \rightarrow o(r^{m_{\mu\nu}})$ for only a subset $\mathcal{S}$ of the indices ${\mu,\nu}$, and the weaker condition $\lie{\xi}{\back} \rightarrow o(r^{m_{\mu\nu}+1})$ on the remaining ${\mu,\nu}$. We call these {\it relaxed asymptotic Killing equations}. We then proceed as before, determining fall-offs for $\Eeq$ and $\fluc$ so that Eq.(\ref{eq-BBCond}) is satisfied, and requiring that the candidate asymptotic symmetries preserve the boundary conditions. After trying this for several subsets $\mathcal{S}$, we found that imposing $\lie{\xi}{\met} = O(\fluc)$ always reduced the set of candidate asymptotic symmetries to a finite-dimensional set.

We therefore follow a slightly different procedure in the next section in order to find an infinite-dimensional ASG of Rindler space. Firstly, we solve a set of relaxed asymptotic Killing equations, and unlike before, we do not require that the asymptotic Killing vectors have the same fall-offs as the exact Killing vectors. Next, we find boundary conditions on $\fluc$ such that the charges associated with these vectors are finite. Then we impose consistency by requiring that the candidate asymptotic symmetries preserve these boundary conditions. Finally, we compute the charges corresponding to these symmetries and check by hand that they are finite, conserved, and integrable, imposing further conditions on both the symmetries and $\fluc$ as necessary so that these conditions are satisfied.

\subsection{Relaxed Boundary Conditions}\label{sec-Relax}

In this section we carry out the steps described above to find an infinite-dimensional ASG for Rindler space. (The details of the calculations are given in Appendix \ref{app-Relaxed}.) We should point out that our search is not exhaustive, and there may be other consistent sets of boundary conditions and asymptotic symmetries that lead to a different ASG.

First we define a set of relaxed asymptotic Killing equations by slightly modifying the original set of asymptotic Killing equations in Eq.(\ref{eq-AKE1})-(\ref{eq-AKE3}). We choose to relax the fall-offs of all the components except $vv$, $vr$, and $rr$ by one order, giving the conditions:
\begin{align}
\lie{\xi}{\bar{g}_{vv}} &\rightarrow o(r),\hspace{1cm} \lie{\xi}{\bar{g}_{vr}} \rightarrow 0,\hspace{1cm} \lie{\xi}{\bar{g}_{vy}} \rightarrow o(r^2),\hspace{1cm} \lie{\xi}{\bar{g}_{vz}} \rightarrow o(r^2)\\
\lie{\xi}{\bar{g}_{rr}} &\rightarrow o(1/r),\hspace{0.6cm} \lie{\xi}{\bar{g}_{ry}} \rightarrow o(r),\hspace{0.5cm} \lie{\xi}{\bar{g}_{rz}} \rightarrow o(r)\\
\lie{\xi}{\bar{g}_{yy}} &\rightarrow o(r^2),\hspace{0.85cm} \lie{\xi}{\bar{g}_{yz}} \rightarrow o(r^2),\hspace{0.35cm} \lie{\xi}{\bar{g}_{zz}} \rightarrow o(r^2)
\end{align}
This is a reasonable choice, as the $v$ and $r$ coordinates describe two-dimensional Rindler space, while $y$ and $z$ are merely transverse coordinates. By imposing the strictest conditions on the $v$-$r$ components of the metric, we are trying to preserve the essence of Rindler space. However, note that we are only using these equations as a guide to finding the correct form of $\xi$, and do not necessarily require them to be satisfied at the end of the analysis. What is important is that the final asymptotic Killing vectors should form a well-defined algebra under the Lie bracket, preserve the final boundary conditions on $\fluc$, and correspond to finite, conserved, and integrable charges.

Solving these equations to leading order and requiring that the vectors $\xi$ form a well-defined algebra under the Lie bracket gives the following fall-offs for the candidate asymptotic Killing vectors:
\begin{equation}
\xi^v,\,\,\xi^y,\,\,\xi^z \rightarrow O(r^0),\hspace{1cm}\xi^r \rightarrow O(r).
\end{equation}
We write a general candidate asymptotic Killing vector in the form
\begin{eqnarray}\label{eq-AKERelaxed}
\xi^v &=&V(v,y,z) + \frac{V_1(v,y,z)}{r} + O(1/r^2)\\
\xi^r &=&rR(v,y,z)+R_1(v,y,z) + O(1/r)\nonumber\\
\xi^y &=&Y(v,y,z) + O(1/r)\nonumber\\
\xi^z &=&Z(v,y,z) + O(1/r)\nonumber
\end{eqnarray}
and calculate the corresponding charge $Q_\xi$. Assuming that we can expand $\fluc$ polynomially to at least second order in $r$ as shown in Eq.(\ref{eq-Expansion}), we find the following conditions must be satisfied in order for the charge to be finite:
\begin{eqnarray}\label{eq-BCondRelaxed1}
h_{yy},\,\,h_{zz} &\rightarrow& O(1)\\
h_{ry},\,\,h_{rz} &\rightarrow& O(1/r)\label{eq-BCondRelaxed2}\\
h_{vy},\,\,h_{vz} &\rightarrow& O(1)\label{eq-BCondRelaxed3}
\end{eqnarray}
Furthermore, if these conditions are satisfied, then $R_1$, $V_1$, $Y$, and $Z$ do not contribute to the charges. Thus these components correspond to trivial asymptotic Killing vectors.
We now require that the vectors of the form (\ref{eq-AKERelaxed}) preserve the boundary conditions (\ref{eq-BCondRelaxed1})-(\ref{eq-BCondRelaxed3}), so that
\begin{eqnarray}
\lie{\xi}{g_{yy}},\,\,\lie{\xi}{g_{zz}} &\rightarrow& O(1)\\
\lie{\xi}{g_{ry}},\,\,\lie{\xi}{g_{rz}} &\rightarrow& O(1/r)\\
\lie{\xi}{g_{vy}},\,\,\lie{\xi}{g_{vz}} &\rightarrow& O(1)
\end{eqnarray}
for any $\met$ allowed by the boundary conditions. If we do not impose constraints on the subleading terms of $\xi$, then the $yy$ and $zz$ components of these equations require $h_{yz}=O(1)$. This requires that $\lie{\xi}{g_{yz}}\rightarrow O(1)$, which is satisfied automatically.
The $ry$ and $vy$ components of these equations impose constraints on the form of $V(v,y,z)$ and $R(v,y,z)$. We find that
\begin{align}
\partial_y V &= \partial_z V = 0\\
\partial_y R &= \partial_z R = 0.
\end{align}
Thus the candidate asymptotic Killing vectors have the form
\begin{eqnarray}\label{eq-RelaxedAKV}
\xi^v &=& V(v) + O(1/r)\\
\xi^r &=& rR(v) + O(1)\nonumber\\
\xi^y &=& O(1)\nonumber\\
\xi^z &=& O(1)\nonumber
\end{eqnarray}
The Lie bracket of these vectors is well-defined. At this point we have the following boundary conditions on the metric:
\begin{align}
h_{\mu\nu} &= \left (\begin{array}{cccc} h_{vv} & h_{vr} & O(1) & O(1) \\
h_{rv} = h_{vr} & h_{rr} & O(1/r) & O(1/r)\\
h_{yv} = h_{vy} & h_{yr}=h_{ry} & O(1) & O(1)\\
h_{zv} = h_{vz} & h_{zr}=h_{rz} & h_{zy} = h_{yz} & O(1)
\end{array} \right )
\end{align}
An analysis of the linearized equations of motion given by Eq.(\ref{eq-Eeq}) gives us the following boundary conditions for $\fluc$: 
\begin{align}\label{eq-RelaxedBC1}
h_{\mu\nu} &= \left (\begin{array}{cccc} O(r) & O(1/r) & O(1) & O(1) \\
h_{rv} = h_{vr} & O(1/r^2) & O(1/r) & O(1/r)\\
h_{yv} = h_{vy} & h_{yr}=h_{ry} & O(1) & O(1)\\
h_{zv} = h_{vz} & h_{zr}=h_{rz} & h_{zy} = h_{yz} & O(1)
\end{array} \right )
\end{align}
It is easy to check that the forms of $\xi$ and $\fluc$ determined by (\ref{eq-RelaxedAKV}) and (\ref{eq-RelaxedBC1}) give a consistent set of boundary conditions and candidate asymptotic Killing vectors satisfying $\lie{\xi}{g_{\mu\nu}} = O(h_{\mu\nu})$. Modulo trivial vectors, the asymptotic symmetry algebra is generated by the vectors
\begin{align}
l_n &= e^{inv}\partial_v\\
t_n &= re^{inv}\partial_r
\end{align}
These vectors obey the algebra
\begin{align}\label{eq-ASA}
[l_m, l_n] &= i(n-m)l_{n+m}\\
[t_m, l_n] &= -imt_{m+n}\\
[t_m, t_n] &= 0
\end{align}
The algebra of the candidate asymptotic symmetries is isomorphic to the direct sum of the Virasoro algebra and a current algebra. We now want to compute the charges corresponding to these vectors. Since finiteness and conservation are no longer guaranteed, we will have to check these by hand. As in Section \ref{sec-RedASG} we obtain the charges by integrating over the boundary of a constant $u$ slice. We integrate over a finite box in the $y$-$z$ plane and assume that the only contribution to the integral comes from the part of $\partial\Sigma$ at $r\to\infty$. As the asymptotic Killing vectors do not transform the box in the $y$-$z$ plane, we can use Eq.(\ref{eq-ChargeNullCoords}) to determine the charge.

In order for the charges $\mathcal{L}_n$ to be conserved, and for the Dirac bracket algebra of the charges to give a representation of the asymptotic symmetry algebra, we find that we must impose the following condition on $\fluc$ on-shell:
\begin{eqnarray}
h_{yy}+h_{zz} &=& O(1/r)\label{eq-HarshCond1}
\end{eqnarray}
It is easy to check by direct calculation that $l_n$ and $t_n$ automatically preserve this condition, both on and off-shell. The $\mathcal{L}_n$ charges are given by
\begin{align}
\mathcal{L}_n &:= Q_{l_n}[h,\bar{g}] =-\frac{1}{16\pi}\int_{\partial\Sigma} e^{inv}\bigl [\partial_z h_{vz} + \partial_y h_{vy} \bigr ]
\end{align}
and obey the conservation law
\begin{align}
D_u\mathcal{L}_n &:= \partial_u\mathcal{L}_n + \{\mathcal{L}_n, \mathcal{L}_0\}=0
\end{align}
This is the correct definition of complete $u$-dependence, as $\mathcal{L}_0$ is the generator of translations in the $v$ direction. The Dirac bracket algebra is
\begin{equation}
\{\mathcal{L}_m,\mathcal{L}_n\} = -\{\mathcal{L}_n,\mathcal{L}_m\} = i(n-m)\mathcal{L}_{n+m}
\end{equation}
The $\mathcal{T}_n$ charges are given by:
\begin{align}
\mathcal{T}_n &:= Q_{t_n}[h,\bar{g}] = \frac{1}{32\pi}\int_{\partial\Sigma} re^{inv}\bigl [h_{yy}+h_{zz}+2(\partial_z h_{rz} +\partial_y h_{ry})\bigr ]
\end{align}
The Dirac bracket algebra of the charges satisfies
\begin{equation}
\{\mathcal{T}_m,\mathcal{T}_n\} = 0
\end{equation}
without requiring any further conditions on the metric. A difficulty arises with the algebra $\{\mathcal{L}_n, \mathcal{T}_m\}$, which is not isomorphic to the algebra of the asymptotic symmetries even after antisymmetrization (More details are given in Appendix \ref{app-Relaxed}.) Furthermore, since $\mathcal{L}_0$ generates translations in the $v$-direction this makes it unclear whether we can consistently define conservation of $\mathcal{T}_m$. Therefore, if we want an algebra of charges that gives a representation of the asymptotic symmetry algebra, we have to discard the symmetries $t_m$, leaving us with one copy of the Virasoro algebra. This can be done consistently since the algebra of the symmetries $l_n$ closes.

Expanding $\fluc$ to first order in $r$ as shown in Eq.(\ref{eq-Expansion}) we see that the integrands are linear in the coefficients $\fluc^1$, so asymptotic linearity holds and the charges are integrable. A direct computation shows that the central charge $\mathcal{K}_{\xi,\xi'}$ is trivial for all asymptotic Killing vectors $\xi$. 

In conclusion, we have found a consistent set of asymptotic Killing vectors $\{l_n, t_m\}$ and boundary conditions for $\fluc$ such that we obtain an infinite-dimensional asymptotic symmetry algebra for Rindler space. The algebra of the asymptotic Killing vectors is isomorphic to the direct sum of a Virasoro algebra and a current algebra. The charges corresponding to the asymptotic symmetries are finite. If we demand that the Dirac bracket algebra of the charges gives a representation of the algebra of asymptotic symmetries, then we must first impose certain other conditions on $\fluc$, and discard the symmetries $t_m$, keeping only the $l_n$ and the corresponding charges. The charges $\mathcal{L}_n$ are also conserved (i.e. independent of $u$) given certain constraints on $\fluc$. The algebra does not have a non-trivial central extension.

We note that repeating this analysis with the coordinate $u$ rather than $v$ does not change the results. That is, we obtain the same $\xi$ and boundary conditions on $\fluc$, and therefore the same asymptotic symmetry algebra as in Eq.(\ref{eq-ASA}). We also obtain the same ASG and Dirac bracket algebra of charges, with the same condition (\ref{eq-HarshCond1}) on the metric fluctuations.

\section{Asymptotic Symmetries of the Rindler Horizon}\label{sec-Horizon}

We now investigate the other boundary of Rindler space, the event horizons at $u=0$ and $r=0$. Since we investigated $\mathcal{J}^+$ by taking $r\to\infty$ when discussing the boundary at infinity, we now study the past horizon by taking the limit $r\to 0$. In order to use the Barnich-Brandt formalism, we make the change of coordinates
\begin{equation}
s = \frac{1}{r},
\end{equation}
so that the metric becomes
\begin{equation}
\mathrm{d}s^2 = -\frac{1}{2a^2 s^2}\mathrm{d}u\mathrm{d}s +\mathrm{d}y^2+\mathrm{d}z^2
\end{equation}
The past horizon is given by the limit $s\to\infty$, with $u$, $y$, and $z$ constant. (We use the coordinate $u$ rather than $v$ for simplicity.)

We make the following ansatz for a general candidate asymptotic Killing vector:
\begin{eqnarray}\label{eq-HorAKERelaxed}
\xi^u &=&U(u,y,z) + \frac{U_1(u,y,z)}{s} + O(1/s^2)\\
\xi^s &=&sS(u,y,z)+S_1(u,y,z) + O(1/s)\nonumber\\
\xi^y &=&Y(u,y,z) + O(1/s)\nonumber\\
\xi^z &=&Z(u,y,z) + O(1/s)\nonumber
\end{eqnarray}
and calculate the corresponding charge $Q_\xi$. We now follow the same strategy as in Section \ref{sec-Relax} to find a consistent set of candidate asymptotic Killing vectors and boundary conditions on $\fluc$. To avoid needless repetition we will simply present the results here. Details of calculations that differ from those in Section \ref{sec-Relax} are given in Appendix \ref{app-Horizon}.

The boundary conditions on the metric are
\begin{align}\label{eq-HorRelaxedBC1}
h_{\mu\nu} &= \left (\begin{array}{cccc} O(1/s) & O(1/s^2) & O(1/s) & O(1/s) \\
h_{su} = h_{us} & O(1/s^3) & O(1/s^2) & O(1/s^2)\\
h_{yu} = h_{uy} & h_{yr}=h_{ry} & O(1) & O(1)\\
h_{zu} = h_{uz} & h_{zr}=h_{rz} & h_{zy} = h_{yz} & O(1)
\end{array} \right )
\end{align}
together with the further constraints
\begin{align}
\partial_u h_{yy},\,\,\, \partial_u h_{zz},\,\,\, \partial_u h_{yz} &\to O(1/s)\\
\partial_u^2 h_{sy} - \partial_s \partial_u h_{uy} &\to O(1/s^3)\\
\partial_u^2 h_{sz} - \partial_s \partial_u h_{uz} &\to O(1/s^3)
\end{align}
These boundary conditions and constraints are preserved by a general candidate asymptotic Killing vector, which (modulo trivial vectors) has the form
\begin{eqnarray}
\xi^u &=&A_1u +O(1/s)\\
\xi^s &=&O(1/s)\nonumber\\
\xi^y &=&Y(y,z) + O(1/s)\nonumber\\
\xi^z &=&Z(y,z) + O(1/s)\nonumber
\end{eqnarray}
for an arbitrary constant $A_1$ and arbitrary functions $Y(y,z)$ and $Z(y,z)$. 

The asymptotic symmetry algebra is generated by the vectors
\begin{align}
y_n^+ &:=e^{in(y+z)}\partial_y\\
y_n^- &:=e^{in(y-z)}\partial_y\\
z_n^+ &:=e^{in(y+z)}\partial_z\\
z_n^- &:=e^{in(z-y)}\partial_z\\
u_0 &:= u\partial_u
\end{align}
These vectors obey the algebra
\begin{align}
[y_n^+,y_m^+] &= i(m-n)y_{n+m}^+,\\
[y_n^-,y_m^-] &= i(m-n)y_{n+m}^-,\\
[y_n^+,y_m^-] &= i(m-n)e^{im(y-z)+in(y+z)}\partial_y,\\
[y_n^+, z_m^+] &=-iny_{n+m}^+ + imz_{n+m}^+\\
[y_n^-, z_m^-] &=iny_{n+m}^- + imz_{n+m}^-\\
[y_n^+, z_m^-] &=e^{im(y-z)+in(y+z)}\left (-in\partial_y + im\partial_z\right )\\
[y_n^{+(-)},u_0] &= 0,
\end{align}
with corresponding relations holding for the $z_n$ vectors. In order to obtain a Dirac bracket algebra of charges that represents the asymptotic symmetry algebra, we discard the vector $u_0$ and only keep $y_n$ and $z_n$ as asymptotic symmetry vectors. The asymptotic symmetry algebra contains more than one copy of the Virasoro algebra. 

The charges corresponding to these vectors are given by
\begin{align}
\mathcal{Y}_n^+ := Q_{y_n^+}[h,\bar{g}] = \frac{a^2}{8\pi}\int_{\partial\Sigma} s^2 e^{in(y+z)} \left [\partial_u h_{sy} - \partial_s h_{uy} \right ]\\
\mathcal{Y}_n^- := Q_{y_n^-}[h,\bar{g}] = \frac{a^2}{8\pi}\int_{\partial\Sigma} s^2 e^{in(y-z)} \left [\partial_u h_{sy} - \partial_s h_{uy} \right ]\\
\mathcal{Z}_n^+ := Q_{z_n^+}[h,\bar{g}] = \frac{a^2}{8\pi}\int_{\partial\Sigma} s^2 e^{in(y+z)} \left [\partial_u h_{sz} - \partial_s h_{uz} \right ]\\
\mathcal{Z}_n^- := Q_{z_n^-}[h,\bar{g}] = \frac{a^2}{8\pi}\int_{\partial\Sigma} s^2 e^{in(y-z)} \left [\partial_u h_{sz} - \partial_s h_{uz} \right ]
\end{align}
They obey the conservation law $\partial_u Q_{\xi} = 0$, and are finite and integrable, assuming that $\fluc$ can be expanded to at least second order in $s$ as shown in Eq.(\ref{eq-Expansion}). The Dirac bracket algebra of the charges correctly represents the asymptotic symmetry algebra, so that
\begin{equation}
\{Q_{\xi_1},Q_{\xi_2}\} = Q_{[\xi_1,\xi_2]}
\end{equation}
Direct computation shows that the algebra does not have a non-trivial central extension.

\section{Discussion}\label{sec-Discussion}

We have investigated the asymptotic symmetries of 4-dimensional Rindler space, both at null infinity and at the past event horizon. Initially we used algorithms that are effective in finding the ASG of AdS, flat, and G$\ddot{\mathrm{o}}$del spacetimes, but these methods only gave finite dimensional ASGs for Rindler space at $\mathcal{J}^+$. Comp$\grave{\mathrm{e}}$re et al. claim in \cite{BBSchrod} that when solving the asymptotic Killing equations to find candidate asymptotic symmetries, the equations should be solved to a suitable order (guided by intuition), as solving only to the leading order will yield a very large class of symmetries. This is clearly not the case for Rindler space, as even solving the asymptotic Killing equations to first order gives a finite-dimensional set of candidate symmetries. Moreover, using the Barnich-Brandt formalism to compute the charges corresponding to these symmetries only gave trivial charges.

We then used relaxed asymptotic Killing equations to find an infinite-dimensional algebra of asymptotic symmetries at $\mathcal{J}^+$, and determined consistent boundary conditions on the metric perturbations simply by imposing finiteness of the charges and studying the linearized equations of motion. We found that the resulting algebra was isomorphic to a direct sum of the Virasoro algebra and a current algebra. We also applied the Barnich-Brandt formalism to the other part of the boundary of Rindler space: namely, the event horizon. We determined a set of consistent boundary conditions and asymptotic Killing vectors, and found that the asymptotic symmetry algebra contained more than one copy of the Virasoro algebra. 

We computed the charges corresponding to the symmetries at $\mathcal{J}^+$ and at the event horizon, and found that they were finite and integrable. If certain constraints were obeyed by the metric, then the charges were also conserved on-shell and their Dirac bracket algebra gave a representation of the asymptotic symmetry algebra. These constraints were preserved by the asymptotic symmetries.

Unfortunately, in all three cases the charge algebra did not have a non-trivial central extension. It remains to be seen whether there are other sets of consistent boundary conditions and asymptotic symmetries that will yield a centrally extended algebra of charges.

\section{Acknowledgments}

I would like to thank Dionysios Anninos for introducing me to this problem, and for pointing me towards useful reference papers.

\begin{appendix}

\section{Computing the restricted ASG of Rindler space}\label{app-Red}

The exact Killing vectors $\xi_0$ of the Rindler background are obtained by carrying out a coordinate transformation on the exact Killing vectors of Minkowski space. They have the following form:
\begin{align}
\xi_0^v &= -\frac{b_{[01]}}{a} - e^{av}(c_0 + c_1 + (b_{[02]} + b_{[12]})y + (b_{[03]}+b_{[13]}) z)\\
\xi_0^r &= -b_{[01]}r + a(-c_0 + c_1 + (-b_{[02]} + b_{[12]})y + (-b_{[03]} + b_{[13]})z)\nonumber\\
\xi_0^y &= c_2 + b_{[23]}z+\frac{(b_{[02]}-b_{[12]})e^{-av} - (b_{[02]}+b_{[12]})r}{2a}\nonumber\\
\xi_0^z &= c_3 -b_{[23]}y+ \frac{(b_{[03]}-b_{[13]})e^{-av} - (b_{[03]}+b_{[13]})r}{2a}\nonumber
\end{align}
where $c_i, b_i$ are arbitrary constants and $b_{[ij]} := b_{ij} - b_{ji}$. The components of the exact Killing vectors of the background fall off like
\begin{equation}
\xi_0^v \rightarrow O(1),\hspace{1cm}\xi_0^r,\,\xi_0^y,\,\xi_0^z \rightarrow O(r)
\end{equation}
at infinity. We therefore write the components of a general asymptotic Killing vector in the form:
\begin{eqnarray}
\xi^v &=& V(v,y,z) + o(r^0)\\
\xi^r &=& rR(v,y,z) + o(r)\nonumber\\
\xi^y &=& rY(v,y,z) + o(r)\nonumber\\
\xi^z &=& rZ(v,y,z) + o(r)\nonumber
\end{eqnarray}
With this form of $\xi$, the $rr$ component of the asymptotic Killing equations is automatically satisfied. The remaining asymptotic Killing equations are:
\begin{eqnarray}
\lie{\xi}{\bar{g}_{vv}} &=& -\frac{e^{-av}}{a}r\partial_v R \rightarrow o(r)\\
\lie{\xi}{\bar{g}_{vr}} &=& -\frac{e^{-av}}{2a}(aV-\partial_v V-R) \rightarrow 0\\
\lie{\xi}{\bar{g}_{vy}} &=& -\frac{re^{-av}}{2a}\partial_y R + r\partial_v Y \rightarrow o(r)\\
\lie{\xi}{\bar{g}_{vz}} &=& -\frac{re^{-av}}{2a}\partial_z R + r\partial_v Z\rightarrow o(r)\\
\lie{\xi}{\bar{g}_{ry}} &=& -\frac{e^{-av}}{2a}\partial_y V + Y \rightarrow 0\\
\lie{\xi}{\bar{g}_{rz}} &=& -\frac{e^{-av}}{2a}\partial_z V + Z \rightarrow 0\\
\lie{\xi}{\bar{g}_{yy}} &=& 2r\partial_y Y \rightarrow o(r)\\
\lie{\xi}{\bar{g}_{yz}} &=& r(\partial_z Y + \partial_y Z) \rightarrow o(r)\\
\lie{\xi}{\bar{g}_{zz}} &=& 2r\partial_z Z \rightarrow o(r)
\end{eqnarray}
This imposes
\begin{eqnarray}
Y &=& \frac{e^{-av}\partial_y V}{2a}\\
Z &=& \frac{e^{-av}\partial_z V}{2a}\\
R &=& -\partial_v V+aV
\end{eqnarray}
together with the following differential equations for $V(v,y,z)$:
\begin{eqnarray}
\partial_y^2 V &=& \partial_z^2 V = 0\\
a\partial_v V - \partial_v^2 V &=& 0\\
\partial_y\partial_z V &=& 0\\
\partial_v\partial_y V -a\partial_y V &=& \partial_v\partial_z V -a\partial_z V=0
\end{eqnarray}
Solving these equations gives
\begin{eqnarray}
\xi^v &=& e^{av}(A_1 + A_2z + B_1y + C_1e^{-av}) + o(r^0)\\
\xi^r &=& aC_1r + o(r)\nonumber\\
\xi^y &=& \frac{B_1r}{2a} + o(r)\nonumber\\
\xi^z &=& \frac{A_2r}{2a} + o(r)\nonumber
\end{eqnarray}
for arbitrary constants $A_1, A_2, B_1,$ and $C_1$. Direct calculation shows that candidate asymptotic Killing vectors of the form
\begin{equation}
\xi^v \rightarrow 0,\,\,\,\,\,\,\, \xi^r,\,\xi^y,\,\xi^z \rightarrow o(r)
\end{equation}
automatically satisfy the asymptotic Killing equations to leading order. Therefore, such asymptotic Killing vectors are trivial, and after requiring the algebra of the vectors to be well-defined under the Lie bracket we can write the components of a general candidate asymptotic Killing vector in the form
\begin{eqnarray}
\xi^v &=& e^{av}(A_1 + A_2z + B_1y + C_1e^{-av}) + O(1/r)\\
\xi^r &=& aC_1r + O(1)\nonumber\\
\xi^y &=& \frac{B_1r}{2a} + O(1)\nonumber\\
\xi^z &=& \frac{A_2r}{2a} + O(1)\nonumber
\end{eqnarray}
For a general $\xi$ of this form, we find that the asymptotic behavior of $\lie{\xi}{\back}$ is
\begin{align}
\lie{\xi}{\bar{g}_{vv}} &\rightarrow O(1),\hspace{1.1cm} \lie{\xi}{\bar{g}_{vr}} \rightarrow O(1/r),\hspace{0.5cm} \lie{\xi}{\bar{g}_{vy}} \rightarrow O(1),\hspace{0.5cm} \lie{\xi}{\bar{g}_{vz}} \rightarrow O(1)\\
\lie{\xi}{\bar{g}_{rr}} &\rightarrow O(1/r^2),\hspace{0.5cm} \lie{\xi}{\bar{g}_{ry}} \rightarrow O(1/r),\hspace{0.5cm} \lie{\xi}{\bar{g}_{rz}} \rightarrow O(1/r)\\
\lie{\xi}{\bar{g}_{yy}} &\rightarrow O(1),\hspace{1.1cm} \lie{\xi}{\bar{g}_{yz}} \rightarrow O(1),\hspace{0.9cm} \lie{\xi}{\bar{g}_{zz}} \rightarrow O(1)\end{align}
The asymptotic Killing equations give us the required boundary conditions on $\Eeq$ for conserved and finite charges:
\begin{align}
\mathcal{H}^{vv} &\rightarrow O(1/r^2),\hspace{0.5cm}\mathcal{H}^{vr} \rightarrow O(1/r),\hspace{0.5cm}\mathcal{H}^{vy} \rightarrow O(1/r^2),\hspace{0.5cm}\mathcal{H}^{vz} \rightarrow O(1/r^2)\\
\mathcal{H}^{rr} &\rightarrow O(1),\hspace{1.1cm}\mathcal{H}^{ry} \rightarrow O(1/r),\hspace{0.5cm}\mathcal{H}^{rz} \rightarrow O(1/r)\\
\mathcal{H}^{yy} &\rightarrow O(1/r^2),\hspace{0.5cm}\mathcal{H}^{yz} \rightarrow O(1/r^2),\hspace{0.4cm}\mathcal{H}^{zz} \rightarrow O(1/r^2)
\end{align}

Calculting $\Eeq$ in terms of $\fluc$ using Eq.(\ref{eq-Eeq}) gives the corresponding boundary conditions for $\fluc$:
\begin{align}
h_{\mu\nu} &= \left (\begin{array}{cccc} O(1) & O(1/r^2) & O(1/r) & O(1/r) \\
h_{rv} = h_{vr} & O(1/r^2) & O(1/r^2) & O(1/r^2)\\
h_{yv} = h_{vy} & h_{yr}=h_{ry} & O(1/r) & O(1/r)\\
h_{zv} = h_{vz} & h_{zr}=h_{rz} & h_{zy} = h_{yz} & O(1/r)
\end{array} \right )
\end{align}
We now require that the candidate asymptotic Killing vectors satisfy $\lie{\xi}{\met} = O(\fluc)$. This places further constraints on the form of $\xi$, including the subleading terms. We therefore write the components of $\xi$ as:
\begin{eqnarray}
\xi^v &=& e^{av}(A_1 + A_2z + B_1y + C_1e^{-av}) + \frac{V_1(v,y,z)}{r}+O(1/r^2)\\
\xi^r &=& aC_1r + R_1(v,y,z) + O(1/r)\nonumber\\
\xi^y &=& \frac{B_1r}{2a}+ Y_1(v,y,z)+O(1/r)\nonumber\\
\xi^z &=& \frac{A_2r}{2a}+ Z_1(v,y,z)+O(1/r)\nonumber
\end{eqnarray}
for arbitrary functions $V_1$, $R_1$, $Y_1$, and $Z_1$. Requiring $\lie{\xi}{\met}=O(\fluc)$ leads to the following constraints on the components of $\xi$:
\begin{align}
\partial_v V_1 -aV_1&= \partial_y V_1 = \partial_z V_1 = 0\\
\partial_v R_1 &= \partial_y R_1 = \partial_z R_1 = 0\\
\partial_v Y_1 &= \partial_y Y_1 = \partial_z Y_1 = 0\\
\partial_v Z_1 &= \partial_y Z_1 = \partial_z Z_1 = 0\\
A_2 &= B_1 = 0
\end{align}
We finally find
\begin{eqnarray}
\xi^u &=& e^{av}\left(A_1 + C_1e^{-av} + \frac{V_1}{r}\right)+O(1/r^2)\\
\xi^r &=& aC_1r + R_1 + O(1/r)\nonumber\\
\xi^y &=& Y_1 + O(1/r)\nonumber\\
\xi^z &=& Z_1 + O(1/r)\nonumber
\end{eqnarray}
for arbitrary constants $A_1, C_1, R_1, V_1, Y_1$, and $Z_1$.

\section{Relaxed boundary conditions}\label{app-Relaxed}

The process of solving relaxed asymptotic Killing equations and finding candidate asymptotic symmetries is similar to that described in Appendix \ref{app-Red}, so we will not describe those calculations here. Instead, we concentrate on the computation of charges corresponding to asymptotic symmetries. 

Given candidate asymptotic Killing vectors
\begin{align}
l_n &= e^{inv}\partial_v\\
t_n &= re^{inv}\partial_r
\end{align}
and assuming as before that we can expand $\fluc$ polynomially to at least second order in $r$, we find the following expressions for the charges:
\begin{align}
\mathcal{L}_n &:= Q_{l_n}[h,\bar{g}] =\frac{1}{32\pi}\int_{\partial\Sigma} e^{inv}\bigl [(in-a)(h_{yy}+h_{zz})+2(\partial_z h_{vz} + \partial_y h_{vy} - \partial_v h_{yy} - \partial_v h_{zz} )\bigr ]\\
\mathcal{T}_n &:= Q_{t_n}[h,\bar{g}] = -\frac{1}{32\pi}\int_{\partial\Sigma} e^{inv}\bigl [h_{yy}+h_{zz}+2r(\partial_z h_{rz} +\partial_y h_{ry}) \bigr ]\label{eq-TCharge}
\end{align}
The charges are indeed finite for $\fluc$ obeying the boundary conditions in (\ref{eq-RelaxedBC1}). Moreover the charges are linear in the first-order coefficients $\fluc^1$ in the expansion of $\fluc$, so asymptotic linearity holds and the charges are integrable around $\back$. 

We now compute the Dirac bracket algebra of the charges given by (\ref{eq-Algebra}). The Dirac bracket of the charges $\mathcal{T}_n$ is
\begin{align}\label{eq-TAlg}
\{\mathcal{T}_m,\mathcal{T}_n\} &= \frac{1}{32\pi}\int_{\partial\Sigma}re^{i(m+n)v}\bigl[2(\partial_z h_{rz}+\partial_y h_{ry})+ 2r(\partial_r\partial_zh_{rz}+\partial_r\partial_zh_{ry}) \bigr]
\end{align}
Using our assumption that $\fluc$ can be polynomially expanded to at least second order in $r$, and the boundary conditions on $\fluc$, we find
\begin{align}
r\partial_r\partial_z h_{rz} &= r\biggl[\partial_r\partial_z \left [\frac{h_{rz}^1}{r}+\frac{h_{rz}^2}{r^2}+O(1/r^3)\right]= -\partial_z h_{rz}
\end{align}
It follows that all the terms in the integrand in (\ref{eq-TAlg}) cancel up to $O(1/r^2)$, so that the integral evaluates to zero and the algebra satisfies
\begin{align}
\{\mathcal{T}_m,\mathcal{T}_n\} &= 0.
\end{align}
The algebra of the $\mathcal{L}_n$ charges, however, is slightly more complicated:
\begin{align}
\{\mathcal{L}_m,\mathcal{L}_n\} &= \frac{1}{16\pi}\int_{\partial\Sigma} e^{i(m+n)v}\bigl [in(\partial_z h_{vz} + \partial_y h_{vy})+i(\frac{m}{2}-n+ia)(\partial_v h_{yy} + \partial_v h_{zz})\nonumber\\
&\hspace{2.5cm}+(\partial_v\partial_z h_{vz} + \partial_v\partial_y h_{vy}-\partial_v^2 h_{yy} - \partial_v^2 h_{zz}) \bigr]
\end{align}
We can define the Dirac bracket by antisymmetrizing
\begin{equation}\label{eq-AntiSymm}
\{Q_{\xi_1}, Q_{\xi_2}\} := \delta_{\xi_2}Q_{\xi_1}-\delta_{\xi_1}Q_{\xi_2}
\end{equation}
to obtain
\begin{equation}
\{\mathcal{L}_m,\mathcal{L}_n\} = \frac{1}{32\pi}\int_{\partial\Sigma}i(n-m)e^{i(m+n)v}\bigl [2(\partial_z h_{vz} + \partial_y h_{vy})-3(\partial_v h_{yy} + \partial_v h_{zz})\bigr]
\end{equation}
Thus if $\fluc$ satisfies
\begin{equation}\label{eq-Assump1}
h_{yy} + h_{zz} = O(1/r)
\end{equation}
on-shell, then 
\begin{equation}
\{\mathcal{L}_m,\mathcal{L}_n\} = -\{\mathcal{L}_n,\mathcal{L}_m\} = i(n-m)\mathcal{L}_{n+m}
\end{equation}
and the Dirac bracket algebra of the charges $\mathcal{L}_n$ represents the algebra of asymptotic Killing vectors $l_n$.
Now we turn to the Dirac bracket of $\mathcal{T}_m$ and $\mathcal{L}_n$. Keeping our assumption (\ref{eq-Assump1}) on the form of $\fluc$, we find
\begin{align}
\{\mathcal{L}_n,\mathcal{T}_m\} &=\frac{i}{16\pi}\int_{\partial\Sigma} mre^{i(m+n)v}(\partial_z h_{rz} + \partial_y h_{ry})\\
\{\mathcal{T}_m,\mathcal{L}_n\} &=-\frac{1}{32\pi}\int_{\partial\Sigma} e^{i(m+n)v}\bigl[2r(\partial_v\partial_z h_{rz} + \partial_v\partial_y h_{ry})\bigr]
\end{align}
Comparing this with the form of $\mathcal{T}_{m}$ in (\ref{eq-TCharge}), we see that the Dirac brackets $\{\mathcal{L}_n, \mathcal{T}_m\}$ and $\{\mathcal{T}_m, \mathcal{L}_n\}$ give unexpected results. Therefore, if we want a well-defined algebra of charges then we can keep either the charges $\mathcal{L}_n$ or the charges $\mathcal{T}_m$, but not both. If we want to keep the charges $\mathcal{L}_n$ then we need to impose the condition (\ref{eq-Assump1}). If we want to keep the charges $\mathcal{T}_m$ then no extra conditions on the metric are needed.

Lastly, we check conservation of the charges. i.e. independence of the coordinate $v$. Note that $\mathcal{L}_0$ generates translations in the $v$ direction. The Dirac bracket $\{\mathcal{L}_n, \mathcal{L}_0\}$ gives the variation in $\mathcal{L}_n$ due to the change in $\met$ under the diffeomorphism $l_0$. As $l_n$ is explicitly dependent on $v$, when calculating the total derivative of $\mathcal{L}_n$ we also need to take into account the variation in $\mathcal{L}_n$ due to the change in $l_n$ under the diffeomorphism $l_0$. Since $\mathcal{L}_{l_0}l_n = \partial_v l_n$, we find
\begin{align}
D\mathcal{L}_n &:= Q_{\partial_v l_n}[h,\bar{g}] + \{\mathcal{L}_n, \mathcal{L}_0\}\\
&= in\mathcal{L}_n-in\mathcal{L}_n\\
&=0.
\end{align}
As the charge $\mathcal{L}_0$ generates translations in $v$, and we cannot keep both the charges $\mathcal{L}_n$ and $\mathcal{T}_n$, we cannot consistently define conservation of the charges $\mathcal{T}_m$. Thus, we keep the symmetries $l_n$ and the corresponding charges $\mathcal{L}_n$, and discard the symmetries $t_n$.

\section{Asymptotic Symmetries of the Rindler Horizon}\label{app-Horizon}

Here we present the details of the calculations used to find the ASG of the Rindler horizon that differ from those used in Section \ref{sec-Relax}, when finding the ASG of Rindler space at null infinity.

The only significant change is in the definition of the Dirac bracket. In Section \ref{sec-Relax}, the asymptotic Killing vectors did not change the shape of the box in the $y$-$z$ plane over which we integrated to obtain the charges. This is no longer the case when studying the ASG of the Rindler horizon, since the asymptotic Killing vectors have non-zero $y$ and $z$ components. We therefore have to use Eq. (\ref{eq-ModBracket}) to define the Dirac bracket of charges.

For example, when computing the bracket $\{Q_{y_n^+}, Q_{y_m^+}\}$, we find that
\begin{eqnarray}
\{Q_{y_n^+}^{\mathrm{box}}[h,\bar{g}], Q_{y_m^+}^{\mathrm{box}}[h,\bar{g}]\} &:=& \delta_{y_m^+}^{h}Q_{y_n^+}^{\mathrm{box}}[h,\bar{g}] + \delta_{\xi_2}^{\mathrm{box}}Q_{y_n^+}^{\mathrm{box}}[h,\bar{g}]
\end{eqnarray}
The first term is the usual Dirac bracket given by (\ref{eq-Algebra}),
\begin{equation}
\delta_{y_m^+}^{h}Q_{y_n^+}^{\mathrm{box}}[h,\bar{g}] = \frac{a^2}{8\pi}\int_{\partial\Sigma} s^2 e^{i(m+n)(y+z)} \left [ im (\partial_s h_{uy} - \partial_u h_{sy}) + \partial_s\partial_y h_{uy} - \partial_u\partial_y h_{sy} \right ]
\end{equation}
and the second term is given by
\begin{align}
\delta_{y_m^+}^{\mathrm{box}}Q_{y_n^+}^{\mathrm{box}}[h,\bar{g}] &:= \int_{\partial\Sigma}  \mathcal{L}_{y_m^+} k_{y_n^+}^{[us]}[h,\bar{g}]\\
&= \int_{\partial\Sigma} e^{im(y+z)}\partial_yk_{y_n^+}\\
&= -\frac{a^2}{8\pi}\int_{\partial\Sigma} s^2 e^{i(m+n)(y+z)} \left [in(\partial_s h_{uy} - \partial_u h_{sy})+ \partial_s\partial_y h_{uy} - \partial_u\partial_y h_{sy} \right ]
\end{align}
We therefore find
\begin{equation}
\{Q_{y_n^+}^{\mathrm{box}}, Q_{y_m^+}^{\mathrm{box}}\} = \frac{a^2}{8\pi}\int_{\partial\Sigma} i(m-n) s^2 e^{i(m+n)(y+z)}  (\partial_s h_{uy} - \partial_u h_{sy})
\end{equation}
We now define the Dirac bracket by antisymmetrizing
\begin{equation}
\{Q_{\xi_1}, Q_{\xi_2}\} := \frac{1}{2}[\delta_{\xi_2}Q_{\xi_1} - \delta_{\xi_1}Q_{\xi_2}]
\end{equation}
which gives
\begin{align}
\{Q_{y_n^+}^{\mathrm{box}}, Q_{y_m^+}^{\mathrm{box}}\} &= \frac{a^2}{8\pi}\int_{\partial\Sigma} i(m-n) s^2 e^{i(m+n)(y+z)}  (\partial_s h_{uy} - \partial_u h_{sy})\\
&= i(m-n)Q_{y_{n+m}^+}
\end{align}
Note that this definition of the antisymmetrized Dirac bracket differs by a factor of $1/2$ from the definition in Eq.(\ref{eq-AntiSymm}). Using this modified definition of the Dirac bracket, we find that the algebra of charges represents the asymptotic symmetry algebra.

\end{appendix}

\end{document}